\documentclass[a4paper,12pt]{article}
\pdfoutput=1
\usepackage{epsfig}
\usepackage{amssymb}
\usepackage{amsfonts}
\usepackage{amsmath}
\usepackage{euscript}
\usepackage{verbatim}
\usepackage{latexsym}
\usepackage{graphicx}
\usepackage{caption}
\usepackage{float}
\usepackage{subcaption}

\usepackage{listings}

\newif\ifdtup

\catcode`\@=11

\@addtoreset{equation}{section}

\def\@normalsize{\@setsize\normalsize{15pt}\xiipt\@xiipt
\abovedisplayskip 14pt plus3pt minus3pt%
\belowdisplayskip \abovedisplayskip
\abovedisplayshortskip \z@ plus3pt%
\belowdisplayshortskip 7pt plus3.5pt minus0pt}

\def\small{\@setsize\small{13.6pt}\xipt\@xipt
\abovedisplayskip 13pt plus3pt minus3pt%
\belowdisplayskip \abovedisplayskip
\abovedisplayshortskip \z@ plus3pt%
\belowdisplayshortskip 7pt plus3.5pt minus0pt
\def\@listi{\parsep 4.5pt plus 2pt minus 1pt
     \itemsep \parsep
     \topsep 9pt plus 3pt minus 3pt}}

\relax

\catcode`@=12

\catcode`\@=11

\def\section{\@startsection{section}{1}{\z@}{3.5ex plus 1ex minus
   .2ex}{2.3ex plus .2ex}{\large\bf}}

\def\SymBoxes#1#2#3#4{\newdimen\un@t \un@t#3%
\raisebox{#1}{\rule{#2\un@t}{#4}\hskip-#2\un@t
\@tempdimb\un@t \advance\@tempdimb by-#4\@tempcntb#2\relax%
\@whilenum{\@tempcntb>0}\do{
\rule{#4}{\un@t}\hskip\@tempdimb \advance\@tempcntb by\m@ne}%
\hskip-#2\un@t \rule[\un@t]{#2\un@t}{#4}%
\rule[\un@t]{#4}{#4}\hskip-#4
\rule{#4}{\un@t}}\hskip-#4}                

\begin{document}

\newcommand{\beq}{\begin{equation}}
\newcommand{\eeq}{\end{equation}}
\newcommand{\bea}{\begin{eqnarray}}
\newcommand{\eea}{\end{eqnarray}}
\newcommand{\beas}{\begin{eqnarray*}}
\newcommand{\eeas}{\end{eqnarray*}}
\newcommand{\defi}{\stackrel{\rm def}{=}}
\newcommand{\non}{\nonumber}
\newcommand{\bquo}{\begin{quote}}
\newcommand{\enqu}{\end{quote}}
\renewcommand{\(}{\begin{equation}}
\renewcommand{\)}{\end{equation}}
\def \eqn#1#2{\begin{equation}#2\label{#1}\end{equation}}

\def\e{\epsilon}
\def\IZ{{\mathbb Z}}
\def\IR{{\mathbb R}}
\def\IC{{\mathbb C}}
\def\IQ{{\mathbb Q}}
\def\IH{{\mathbb H}}
\def\de{\partial}
\def\Tr{ \hbox{\rm Tr}}
\def\H{ \hbox{\rm H}}
\def\HE{ \hbox{$\rm H^{even}$}}
\def\HO{ \hbox{$\rm H^{odd}$}}
\def\K{ \hbox{\rm K}}
\def\Im{ \hbox{\rm Im}}
\def\Ker{ \hbox{\rm Ker}}
\def\const{\hbox {\rm const.}}
\def\o{\over}
\def\im{\hbox{\rm Im}}
\def\re{\hbox{\rm Re}}
\def\bra{\langle}\def\ket{\rangle}
\def\Arg{\hbox {\rm Arg}}
\def\Re{\hbox {\rm Re}}
\def\Im{\hbox {\rm Im}}
\def\exo{\hbox {\rm exp}}
\def\diag{\hbox{\rm diag}}
\def\longvert{{\rule[-2mm]{0.1mm}{7mm}}\,}
\def\a{\alpha}
\def\dag{{}^{\dagger}}
\def\tq{{\widetilde q}}
\def\p{{}^{\prime}}
\def\W{W}
\def\N{{\cal N}}
\def\hsp{,\hspace{.7cm}}

\def\br{\nonumber}
\def\IZ{{\mathbb Z}}
\def\IR{{\mathbb R}}
\def\IC{{\mathbb C}}
\def\IQ{{\mathbb Q}}
\def\IP{{\mathbb P}}
\def \eqn#1#2{\begin{equation}#2\label{#1}\end{equation}}

\newcommand{\C}{\ensuremath{\mathbb C}}
\newcommand{\Z}{\ensuremath{\mathbb Z}}
\newcommand{\R}{\ensuremath{\mathbb R}}
\newcommand{\rp}{\ensuremath{\mathbb {RP}}}
\newcommand{\cp}{\ensuremath{\mathbb {CP}}}
\newcommand{\vac}{\ensuremath{|0\rangle}}
\newcommand{\vact}{\ensuremath{|00\rangle}                    }
\newcommand{\oc}{\ensuremath{\overline{c}}}
\newcommand{\psizero}{\psi_{0}}
\newcommand{\phizero}{\phi_{0}}
\newcommand{\hzero}{h_{0}}
\newcommand{\psiin}{\psi_{\rh}}
\newcommand{\phiin}{\phi_{\rh}}
\newcommand{\hin}{h_{\rh}}
\newcommand{\rh}{r_{h}}
\newcommand{\rb}{r_{b}}
\newcommand{\psibnd}{\psi_{0}^{b}}
\newcommand{\psibndp}{\psi_{1}^{b}}
\newcommand{\phibnd}{\phi_{0}^{b}}
\newcommand{\phibndp}{\phi_{1}^{b}}
\newcommand{\gbnd}{g_{0}^{b}}
\newcommand{\hbnd}{h_{0}^{b}}
\newcommand{\zh}{z_{h}}
\newcommand{\zb}{z_{b}}
\newcommand{\man}{\mathcal{M}}
\newcommand{\hbr}{\bar{h}}
\newcommand{\tbr}{\bar{t}}

\begin{titlepage}
\begin{flushright}
CHEP XXXXX
\end{flushright}
\bigskip
\def\thefootnote{\fnsymbol{footnote}}

\begin{center}
{\Large
{\bf Critical Islands 
}
}
\end{center}

\bigskip
\begin{center}
Chethan KRISHNAN$^a$\footnote{\texttt{chethan.krishnan@gmail.com}}  
\vspace{0.1in}

\end{center}

\renewcommand{\thefootnote}{\arabic{footnote}}

\begin{center}

$^a$ {Center for High Energy Physics,\\
Indian Institute of Science, Bangalore 560012, India}\\

\end{center}

\noindent
\begin{center} {\bf Abstract} \end{center}

We discuss a doubly-holographic prescription for black holes in braneworlds with a vanishing cosmological constant. It involves calculating Ryu-Takayanagi surfaces in AdS black funnel spacetimes attached to braneworld black holes in the $ critical$ Randall-Sundrum II model. Critical braneworlds have the virtue of having massless gravitons. Our approach should be useful when the braneworld is a cosmological black hole interacting with deconfined, large-$N$ matter. In higher dimensions, explicit funnel metrics will have to be constructed numerically -- but based on the general structure of the geometry, we present a natural guess for where one might find the semi-classical island. In a 3-dimensional example where a toy analytic black funnel is known, we can check our guess by direct calculation. We argue that this resolves a version of the information paradox in these braneworld systems, by finding strong evidence for ``cosmological islands". Comoving Ryu-Takayanagi surfaces and associated UV cut-offs on the brane, play natural roles.

\vspace{1.6 cm}
\vfill

\end{titlepage}

\setcounter{footnote}{0}

\section{Introduction}

Understanding the Page curve is believed to be a key step in resolving the black hole information paradox \cite{Hawking, Page, Mathur, AMPS}. Recent developments \cite{Penington, Almheiri} suggest that entanglement wedge \cite{EW} phase transitions and associated {\em islands} may be the dynamical mechanism behind the Page curve, when an AdS black hole is allowed to evaporate via a coupling to an external sink. In \cite{Mahajan1, Mahajan3} a further proposal for encoding islands in terms of classical Ryu-Takayanagi surfaces \cite{RT, HRT} in a higher dimensional geometry  was made. To distinguish this from the original proposal in \cite{Penington, Almheiri}, in this paper we will refer to \cite{Mahajan1, Mahajan3} as the {\em  semi-classical island} prescription\footnote{We are following the title of \cite{Mahajan1} in this terminology. We hope this is not too confusing, because ultimately the prescription of \cite{Penington, Almheiri} is also based on semi-classical ingredients \cite{EW}.}. 

\begin{figure}[H]\centering
	\hspace{-5mm}
	\includegraphics[height=90mm,width=100mm,trim = {0 0cm 0 10cm}, clip]{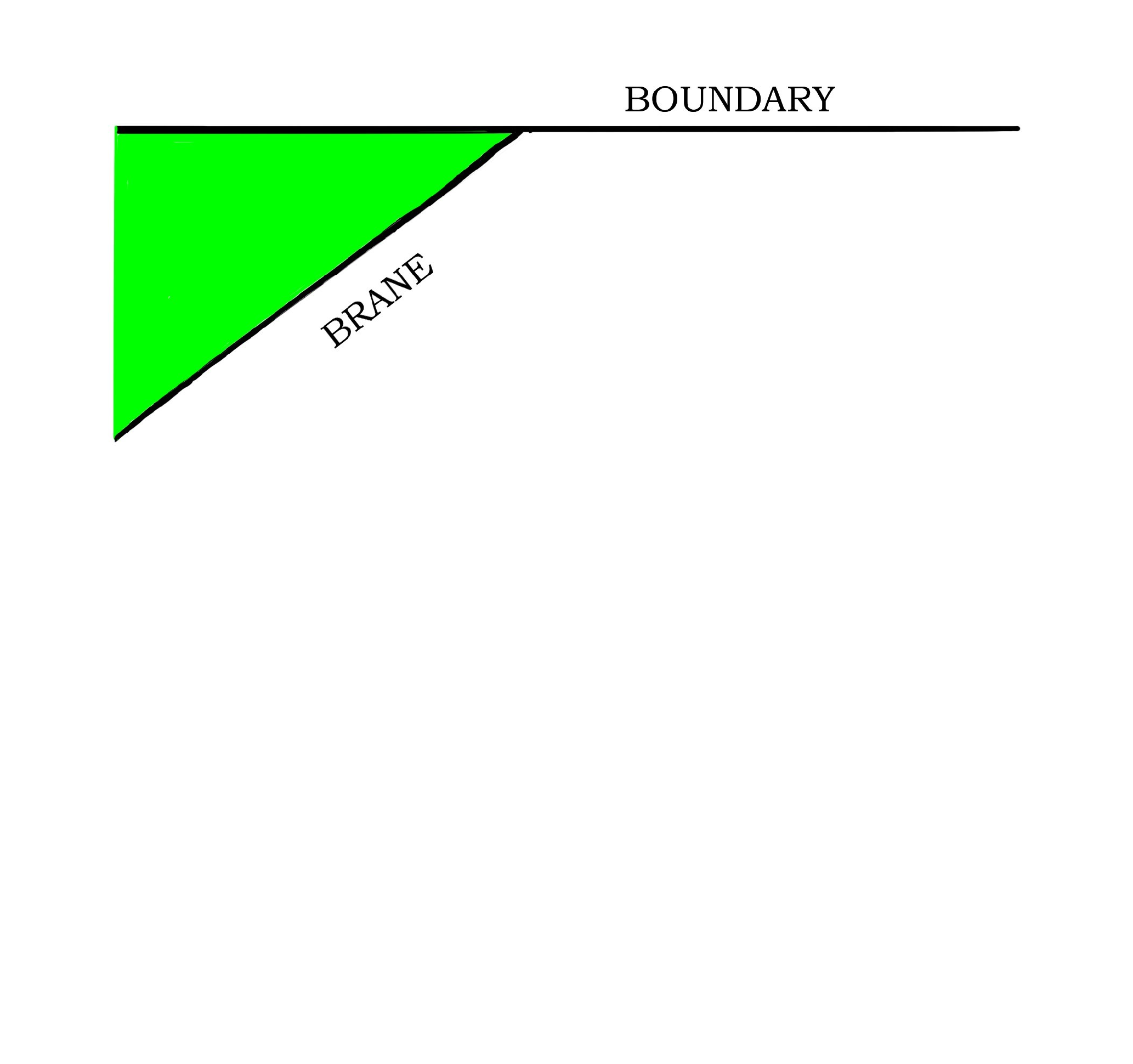} \\
	\vspace{-5cm}
	\caption{The sub-critical Karch-Randall braneworld.}
	\label{subcrit}
\end{figure}

A key point about the semi-classical island prescription is that it is ``doubly-holographic". It uses an auxiliary higher dimensional AdS geometry to capture the entanglement structure of a gravity theory coupled to matter. In other words, this is an example of a braneworld idea. Specifically, it is a braneworld set up where the brane is {\em sub-critical} \cite{KR}. Such a brane reaches the boundary of the bulk, and gravity on it is AdS gravity. A cartoon for this set up is given in Fig. \ref{subcrit}.  
\noindent We will call this the {\em Karch-Randall braneworld} scenario, to be distinguished from the more familiar {\em Randall-Sundrum braneworlds} of \cite{RSII}. The latter are {\em critical} in the sense that gravity on them has vanishing cosmological constant.

A second related point about the semi-classical island prescription of \cite{Mahajan1, Mahajan3} is that it demands that the matter theory (in the matter+gravity system on the brane) be holographic. So it is natural to take the matter to be strongly coupled with large central charge. Equally significantly, for the Hawking radiation on the brane to be visible semi-classically in the bulk\footnote{These nuances are not too significant in the 1+1 dimensional setting, which is the context of \cite{Mahajan1, Mahajan2}.}, the allowed radiation must contain color {\em non-singlets} \cite{Liam}. We will take this to mean that the entanglement wedge/QES arguments of \cite{Penington, Almheiri} are the natural way to think of islands when the brane matter is a weakly coupled non-holographic theory, while the doubly holographic formulation of \cite{Mahajan1, Mahajan3} is the natural language when the brane matter is a deconfined large-$N$ theory. This suggests that the simplest higher dimensional context where the semi-classical island prescription will have a clean formulation is in an eternal black hole above the deconfinement temperature\footnote{It may be possible to model an evaporating black hole also in this context by considering non-equilibrium situations above the deconfinement temperature. A natural setting for this could be the black droplets of \cite{Hubeny2, Santos2}, but we will not say anything about them in this paper. 
}.
When the black hole is held in equilibrium with a thermal bath above the deconfinement temperature, there do exist versions of the information paradox \cite{Mathur2, Mahajan2} where the doubly holographic approach can be straightforwardly formulated \cite{Mahajan2, Mahajan3}. This will be the primary context of this paper. 

The above discussions are in AdS. But in \cite{KPP}, evidence was given that analogues of entanglement wedges must exist not just in AdS, but also in flat space (and  not just in 1+1 dimensions, but in arbitrary dimensions). The arguments relied on identifying two key bulk objects  \cite{CK} -- Asymptotic Causal Diamonds (ACD) and holographic screens. The spatial boundary of flat space does not lend itself to a direct adaptation of the notion of a subregion (unlike the AdS boundary), but it turned out that despite this, ACD-inspired ideas lead to natural definitions of QES and bulk entanglement wedges. This set up clarifies the parallel between flat space information paradox, and the previously considered AdS-plus-sink problems of \cite{Penington, Almheiri}. See also \cite{Mahajan1, Mahajan2, Raju, Thorlacius, Iizuka, Strominger, Ali, Junggi, LCZ} for various related perspectives.  

The arguments of \cite{KPP} gave evidence for entanglement phase transitions in flat space, parallel to the AdS arguments of \cite{Penington}. However, the analogues of the doubly holographic picture of \cite{Mahajan1, Mahajan3} were not discussed. We wish to take a step towards filling this gap here. Identifying a doubly holographic context where braneworld islands emerge, will be complementary evidence for the island idea away from AdS. In this paper, we will present a picture  based on the critical Randall-Sundrum II model \cite{RSII} instead of the sub-critical Karch-Randall: the brane never reaches the conformal boundary. We will argue that  black funnels \cite{Hubeny} anchored to braneworld black holes lead to a natural geometrization, and that holographic entanglement entropy computed in black funnel geometries give evidence for the existence of islands. The black funnels we are interested in are naturally associated to deconfined Hartle-Hawking states of the dual large-$N$ gauge theory \cite{Hubeny}, so this also ties in with our previously stated expectation about the context where double-holography applies. Because funnels are asymptotically planar black holes in the boundary directions, one expects these braneworlds to be cosmologies \cite{Kraus}. 

The main goal of the paper is to make it plausible that there exists doubly holographic contexts where versions of the information paradox can be phrased and resolved in situations without a cosmological constant. This gives a complementary perspective on the results of \cite{KPP}, in situations where the black hole is in a cosmology and interacting with strongly coupled deconfined matter. While our arguments and observations are expected to  be general, due to the scarcity of analytically tractable black funnel metrics, our detailed checks will be limited to a toy AdS$_3$ black funnel in AdS/CFT. In higher dimensions, we expect similar conclusions because they are largely structural -- but there, calculations will have to be numerical. Note that even the AdS/CFT black funnels in higher dimensions are known only numerically \cite{Santos-Way}. So we should emphasize that the braneworld funnels that we are after have not been constructed in the literature -- even though we believe our arguments for their existence are quite plausible.  There are two reasons why we think the scenario we present is reasonable. Firstly, we show by direct calculation in AdS$_3$ funnel spacetimes that there  exist horizon-straddling extremal surfaces of the type needed for islands to be possible. These, need not have existed. Secondly, we will see that the cosmological scaling of the co-moving entanglement entropy on the brane is precisely what one would need for the island argument to work. So despite the speculative nature of some of our arguments, we believe that the picture we present is robust. See a closely related discussion that has appeared in the literature since our work, \cite{Hartman0}.

Before we proceed, let us emphasize a conceptual question regarding the information paradox in asymptotically flat space. We first consider the case of an {\em evaporating} black hole to set the stage, even though that will not be our focus in this paper. In order to unambiguously define a Page curve, one needs to have a clear separation (ie., tensor factorization) between the black hole Hilbert space and the sink Hilbert space. Unlike AdS coupled to an explicit sink, in the asymptotically flat case, one might worry that the dynamical nature of gravity in the asymptotic region might invalidate this separation. We believe this is not a serious issue. The key point is that for an evaporating black hole, (a) the backreaction in the asymptotic region can be made arbitrarily small, and (b) the propagation time for the radiation from a suitablly large cut-off surface to the asymptotic boundary is infinite. Therefore it is reasonable to treat the asymptotic region as a sink for the radiation\footnote{Note that even though the backreaction near the boundary can be made arbitrarily small in AdS, the propagation time to the boundary is finite. So one needs to explicitly couple the system to a sink, to avoid refelection of the radiation back into the black hole (if the black hole is large).}. This and a few other plausibility arguments were made in \cite{CK, KPP} to suggest that at least for some questions, a radial cut-off  is a meaningful and useful object to consider in flat space. In particular, \cite{CK, KPP} noticed that many questions related to entanglement can be given natural flat space analogues of AdS discussions (including quantum error correction), if one works with ACDs together with radial cut-offs. 

This presumed tensor factorization between the black hole system and the sink system is equivalent to the adoption of the so-called ``central dogma", canonized in \cite{Malda-review}. This is the idea that one should treat the black hole as an ordinary thermal system when seen from sufficiently far outside, even in flat space. A further set of arguments, more Euclidean in flavor, were made in \cite{Dong0}. They note that one can define entropy for a gravitating system (which they call effective enrtopy) by introducing cut-off branes in Euclidean geometries and working with path integrals. This has parallels to the argument of Gibbons-Hawking \cite{GH} where a cut-off is introduced in Eucliden signature to do a background subtraction when computing entropies and actions. The arguments in \cite{CK, KPP} were made directly in Lorentzian signature, because the information paradox is a Lorentzian problem.

A potentially more serious issue exists when one is considering the {\em eternal} black hole version of the information paradox in flat space. In order to formulate the problem, we need to couple the black hole to a heat/radiation bath to which it is in equilibrium, in the asymptotic region. Since gravity is dynamical, this means that one {\em cannot} neglect the backreaction in the asymptotic region. In some of the discussions of the Page curve for the flat space eterenal black hole in the literature, this backreaction is simply ignored. A more natural thing in our view\footnote{Note for example, that the Hartle-Hawking state of the flat space eternal black hole represents the black hole in {\em unstable} thermal equilibrium with its own radiation \cite{Kay}.} is to treat the radiation bath as approximately homogeneous and isotropic in the asymptotic region and confront the fact the system is now a black hole in an FRW cosmology supported by radiation. Remarkably, we will see that the doubly holographic set up automatically knows about this! There is a natural understanding of the cosmological nature of the system because the brane is moving in the bulk. 

We will find natural prescriptions to holographically compute entanglement entropies in such cosmological braneworlds. Holographic entanglement entropy is manifestly finite on braneworlds, and it should therefore be understood from the brane point of view as coming pre-equipped with a UV cut-off. We will see how the entanglement entropy with a fixed physical UV cut-off is related to that with a fixed comoving UV cut-off, and how the latter is what keeps track of black hole information. We will also see how this fact plays a role in making the island argument (doubly) holographically.

As a corollary of our discussions, we will be able to make some comments about a recent question \cite{Geng} about graviton masses in these set ups. Even though it never came up in the 1-brane JT-gravity discussions in \cite{Mahajan1}, gravitons in the Karch-Randall braneworld are well-known to be massive in higher dimensions \cite{KR, Porrati}. It was recently observed \cite{Geng}, that in the limit that the mass of these gravitons go to zero, the islands also stop contributing to the physics. If one reads this result to mean that massive gravitons are essential to the success of the island prescription, one might worry whether islands are of any significance at all for the ``real world" information paradox. Our set up sheds some light on this. Massive gravitons arise in Karch-Randall because of the leaky boundary conditions imposed for gravitons in AdS coupled to the sink CFT, and the associated non-conservation of the stress tensor \cite{Porrati}. The limit where the graviton mass goes to zero in the Karch-Randall setting, is also the limit where brane gravity turns off and the brane cosmological constant goes to zero - there is only a single parameter (the angle of the brane) controlling Newton's constant, the braneworld cosmological constant and the graviton mass. The underlying reason for this is that in this limit, the entire brane goes to the boundary and gravity on it becomes non-dynamical.  On the other hand, in the critical RSII setting the graviton is massless and the brane cosmological constant is zero to start with, but as long as we are at finite cut-off in AdS, the $G_N$ is finite. We are no longer trying to get flat space information paradox as a limit of the AdS+sink information paradox, by tuning a single parameter that controls multiple relevant quantities. Such a possibility was indeed envisaged in \cite{Geng}. Note also that as long as we are working with leaky AdS gravity, it is not unreasonable (and indeed is expected \cite{Porrati}) that the graviton is massive.  The key difference between the discussions of \cite{Mahajan1, Mahajan3, Mahajan2, Geng} and what we present here is that we are not aiming to reproduce an AdS braneworld. We view massive gravitons as a feature of AdS braneworlds when coupled to external CFTs via transparent boundary conditions \cite{KR, Porrati}. Our results further strengthen the case that it is the existence of a suitable approximate tensor factorization where one factor contains a black hole, that is key for finding islands and the Page curve. 

\section{Black Holes on RSII Braneworlds}

As we mentioned earlier, the simplest context where we can hope to argue for the existence of semi-classical islands is when the braneworld gravity is coupled to holographic (ie., strongly coupled, large central charge) matter that is deconfined. In other words we need a braneworld black hole that is interacting with deconfined plasma. There exists versions of the information paradox that can be phrased \cite{Mathur2} and resolved \cite{Mahajan2, Mahajan3} for eternal black holes, and we hope to adapt a version of it here. The trouble of course is that no-one has so far managed to construct a strongly coupled braneworld black hole of this type. 

But all is not lost, because there do exist some closely related constructions in the literature. We will string these together to present a picture, where semi-classical islands play a natural role. To proceed, we will use the following chain of observations:

\begin{itemize}
\item AdS geometries where black holes on the asymptotic boundary are interacting with deconfined plasma are known, eg., \cite{Hubeny, Santos-Way}. Note that boundary black holes are fixed backgrounds with non-dynamical gravity, and so ultimately we are interested in finite cut-off braneworld versions of these solutions which need to be explicitly constructed. But they provide useful intuition. In the bulk, these solutions extend to two broad categories of black solutions - black funnels and black droplets. 
\item The black funnel solutions are believed to be duals of the strongly coupled Hartle-Hawking state, ie., they are dual to the deconfined phase of the boundary field theory in equilibrium with the boundary black hole. 
\item Analytic funnel solutions are known in AdS$_3$ and AdS$_4$ \cite{Hubeny}. The latter are adaptations of the so-called AdS C-metrics \cite{Plebanski}. Apart from the fact that these are fairly low-dimensional, the latter also suffer from the feature that the boundary black hole lives in an asymptotically hyperbolic slicing and not in asymptotically flat space. We will use the analytic funnel in AdS$_3$ to do some explicit calculations. The fact that they yield pro-island evidence will be taken as evidence for the existence of islands more generally. The motivation for this comes from the next bullet point.
\item Even though analytic solutions are not known, there do exist numerical black funnel solutions in AdS$_5$ \cite{Santos-Way} that are dual to asymptotically flat boundary Schwarzschild black holes. We will take this as conclusive evidence that black funnels are the holographic description of strongly coupled Hartle-Hawking states on the boundary of AdS. Numerical black funnels in AdS$_4$ that asymptote to three dimensional asymptotically flat black holes have also been constructed \cite{Santos-Way}, even though the analytic C-metric results of \cite{Hubeny} were limited to asymptotically hyperbolic black holes. This is comforting for the narrative we are presenting here in favor of flat space islands. Note further that in 2+1 dimensions there are no black holes in vacuum Einstein gravity. So it is reasonable to view them as related to black holes that are supported by (holographic) matter.  
\item So far of course, we have said nothing about braneworlds and strongly coupled black holes on them. In a very interesting paper \cite{Figueras} (see also \cite{Skenderis}), Figueras and Wiseman have shown that braneworld black holes can be constructed from the associated AdS-CFT problem. More precisely, they argued that given an asymptotic boundary metric and the demand that the behavior of the solution matches with AdS at the Poincare horizon, one can construct corresponding braneworld geometries that are solutions of the induced braneworld gravity.
\item Using this method, Schwarzschild black holes that are asymptotically flat (on the brane) have been constructed on RSII braneworlds \cite{Figueras}, starting from numerical AdS geometries \cite{Lucietti} that asymptote to Schwarzschild black holes at the boundary. 
\item These solutions are not quite what we want, because on the boundary they asymptote to the vacuum, and are more analogous to Unruh/Boulware states than Hartle-Hawking states. In the bulk this is tied to the fact that they asymptote to Poincare horizons and not planar black holes. To get corresponding black funnels, we need to have a horizon that asymptotes to the planar black hole horizon in the interior, in the boundary asymptotic directions. 
\item But nonetheless, from these observations, it seems quite plausible to us that there should also exist a braneworld black hole that in the bulk asymptotes to a planar black hole. We will call this a {\em braneworld black funnel}, even though more accurately it is a black funnel attached to the brane world black hole. We will assume that demanding the planar black hole behavior in the interior and using the numerical black funnels of \cite{Santos-Way} that are anchored to the conformal boundary, and following the philosophy and techniques of \cite{Figueras} such a solution can be constructed. Assuming it exists, this task will at least be of as much difficulty as the construction of the numerical braneworld black hole in \cite{Figueras}. One key difference we expect from \cite{Figueras} is that we expect these black holes to live in braneworld cosmologies, because braneworlds in planar AdS black holes are FRW cosmologies \cite{Kraus}.
\end{itemize}

To summarize -- we have not been able to find explicit braneworld black funnels in the literature, which can be viewed as the strongly coupled black holes that we are after. But the arguments above indicate that braneworld black funnels are closely related to the AdS/CFT black funnels, which have been constructed numerically in some interesting cases even in high enough dimensions \cite{Santos-Way}. So in what follows, we will present a doubly-holographic prescription motivated by \cite{Mahajan1, Mahajan2} assuming that such braneworld black funnel constructions are possible. We will be able to do some concrete checks of our prescription using the toy analytic black funnel constructed in AdS$_3$ by \cite{Hubeny}. 

\section{Critical Islands from the Geometry of Black Funnels}



In this section, we will discuss the structure of the funnel geometry and identify the candidate location where we expect the island contribution to arise. The explicit metric will not be necessary to make our arguments, general properties will suffice. This is good for two reasons. Firstly, as we mentioned, in higher dimensions the physically most interesting black funnels are only known numerically. Secondly, since it is the structure of the geometry that affects the existence of the extremal surface (and not directly the presence or absence of propagating gravitons), it is reasonable to think that the calculation in the AdS$_3$ funnel that we will do in the next section captures the relevant physics even in higher dimensions.  This is loosely analogous to how \cite{Mahajan3} generalized \cite{Mahajan2}.

The geometry in \cite{Mahajan3} is such that there exists an extremal surface anchored at the boundary, that reaches the sub-critical brane in Fig. \ref{subcrit}. We will  call this the {\em second} extremal surface in our discussions below. When the bulk has a planar AdS black hole horizon, there is also an extremal surface that passes through the planar horizon which we call the {\em first} extremal surface. Illustrations of both these extremal surfaces can be found, say, in figure 1 of \cite{Geng}. The competition between these two extremal surfaces is what leads to a phase transition in the entropy curve of the eternal black hole at late times. In more detail, the extremal surface that passes through the planar horizon starts out at $t=0$, as the smaller of the two. So the entanglement entropy is determined by the first extremal surface at early times. But as time progresses, we expect this area to increase linearly with time -- this is because nice slices inside horizons get stretched with time because the black hole interior is a time-dependent background \cite{Hartman}. Since the area of the second extremal surface stays fixed, this means that at late times, the second extremal surface is what determines the RT surface. This is the origin of islands. 

Our goal is to identify parallel structures for the critical brane, and our claim is that black funnel geometries will help us do that. To begin with, let us note that the black funnel is an AdS geometry where the horizon has a neck region that reaches the boundary, see eg. fig 1 of \cite{Hubeny}, or our Fig. \ref{funnel}. In AdS$_3$ the neck has collapsed (see Fig. \ref{AdS3}), but the structure is still essentially the same. Clearly, the first kind of extremal surfaces that are anchored to the asymptotic regions of the boundary and fall through the planar horizon, should still exist. The question is, what replaces the second class of extremal surfaces. It cannot be the case that these extremal surfaces cut the neck of the funnel. Because if they do, by the argument we made above \cite{Hartman}, their areas will keep increasing quasi-linearly with time. This would mean that no matter which extremal surface we picked, the late-time entropy is relentlessly increasing -- there is no saturation of the entropy as one expects in physical systems, and we are left with a version of the information paradox. Cutting the horizon neck has the further and more obvious problem that it does not lead to an island on the braneworld\footnote{But perhaps in some related context, such a scenario could give rise to the intriguing possibility of {\em submerged} islands. Note that a surface that cuts the horizon presumably has to end on a different asymptotic region (or another braneworld in such an asymptotic region).}.

The natural other possibility is that the island contribution comes from a region on the brane near (and outside) the location where the funnel horizon cuts the braneworld. This would be consistent with the fact that on eternal black holes, the island can be outside the horizon, see \cite{Mahajan2}. The relevant RT surface area has a chance of remaining constant as time evolves\footnote{We will discuss how this works out in more detail, in section 5. The story is interesting on its own because it has a braneworld cosmology.}. We will discuss the geometry of this configuration in the following, which will also help our discussion in the next section.


\begin{itemize}
\item A key observation is connected to the discussion in section 2.2 (and the boundary conditions for integration in section 2.3) of \cite{Santos-Way}. In particular see Figure 2 in that paper. From the structure of the black funnel, one would expect the region exterior to the geometry to be characterized by three natural boundaries. This leads to a triangular region (Fig. \ref{triangle}), that arises from the funnel horizon, asymptotic planar black hole and the boundary.  \begin{figure}[H]\centering
	\includegraphics[height=90mm,width=100mm,trim = {0 0cm 0 8cm}, clip]{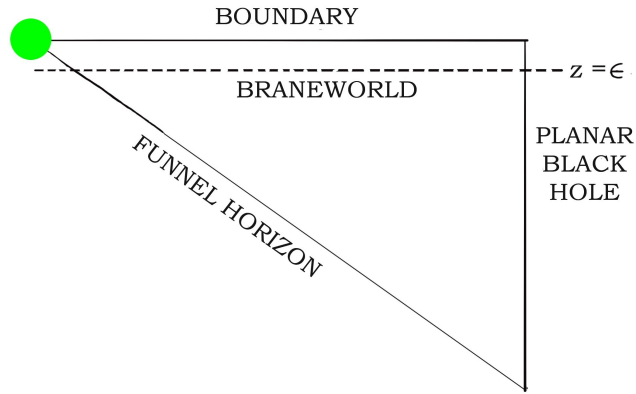} \\
	\vspace{-3cm}
	\caption{A schematic picture of the black funnel. The green dot is where the funnel horizon hits the boundary, $z$ is a holographic coordinate and the horizontal direction is the boundary radial direction.}
	\label{triangle}
\end{figure} 
\noindent 
But it turns out that for a black funnel in dimensions higher than three, one can argue that the region of the horizon where it joins the boundary must look like that of a zero energy hyperbolic black hole. This naturally turns the triangular region of interest in the geometry into a rectangular one: \begin{figure}[H]\centering
	\includegraphics[height=90mm,width=90mm,trim = {0 0cm 0 5cm}, clip]{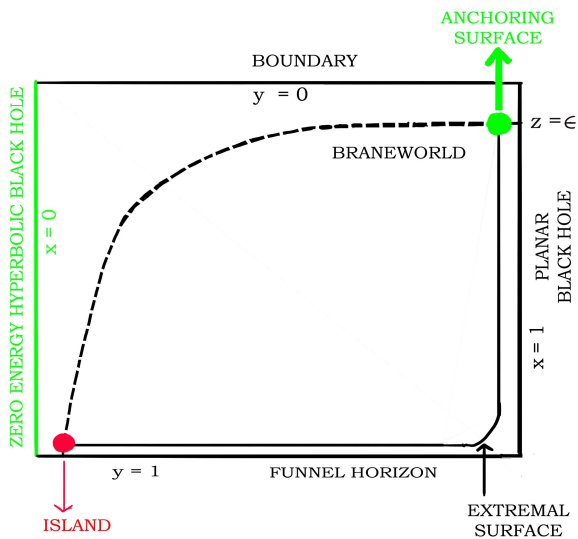} \\
	\vspace{-2cm}
	\caption{A coordinatization of higher dimensional funnels that exhibits their parallels with the 3-dimensional funnels of the next section. }
	\label{rectangle}
\end{figure} 
\noindent
This rectangular configuration makes a heuristic parallel between higher dimensional funnels and the  AdS$_3$ funnel of Fig. \ref{AdS3}  -- instead of the zero energy hyperbolic black hole, in AdS$_3$ we have a collapsed horizon. In the coordinates of \cite{Santos-Way} the rectangle is demarcated by the coordinate boundaries $x=1$ (planar black hole), $y=0$ (the funnel horizon), $y=1$ (boundary of AdS), and $x=0$ (the hyperbolic black hole).  
\item As long as we are away from the funnel horizon ($y=1$ in the notation of section 2.3 of \cite{Santos-Way}) we are outside the horizon of the hyperbolic black hole. The ``neck" region (green line, in figure \ref{rectangle}) of the latter metric is different from the funnel horizon defined by $y=1$. Again this has parallels in AdS$_3$ where the collapsed neck is structurally different from the planar black hole horizon.

\item 
The hyperbolic black hole is an infinite proper distance away from the $x=1$ end of the rectangle. This is because of the stretching due to the hyperbolic metric \cite{Santos-Way}, and is most easily seen by noting that the zero-energy hyperbolic $(d+1)$-black hole metric is approximately 
\bea
d s_{\mathrm{H}}^{2}\approx\frac{L^{2}}{y}\left[-\left(1-y\right) d t^{2}+\frac{d y}{4 y(1-y)}+\frac{d x^{2}}{4x^2}+\frac{1}{4 x}  d \Omega_{d-3}^{2}\right]
\eea
near $x=0$. When the other coordinates are held fixed, we see that the proper distance to $x=0$ diverges. This means that no extremal surface that reaches the hyperbolic black hole can be the second extremal surface. It will never be smaller in area than the first (ie., the one that cuts the planar black hole). Note also that because it is a divergent quantity, this latter statement is true, no matter how long we wait.   
\item Fortunately, this is not quite the final story because so far we have only been dealing with the AdS/CFT geometry, we have not taken the brane in the RSII braneworld into consideration. The braneworld black hole is supposed to live on a brane that is at the coordinate $ z =\epsilon$, where $z$ is the radial coordinate in the Fefferman-Graham expansion. Since the braneworld is time-dependent (ie., the brane must move, and therefore the geometry on it is time-dependent) this should be viewed as a statement at a particular time. What we call $z$ was called $\tilde z$ in the notations of \cite{Santos-Way}. Fefferman-Graham coordinates are the relevant ones for placing the brane in the language of \cite{Skenderis, Figueras} in a Randall-Sundrum II set up. 
\item Interestingly, the coordinates $x, y$ and $z$ are related via
\bea
y x (1+ x)^{2}=z^{2}
\eea
when either $x$ or $y$ is small, see section 3.1 of \cite{Santos-Way}. This means that the location of the brane in the $x$-$y$ coordinate rectangle is schematically as shown in the dashed line of Fig. \ref{rectangle}. In particular, note that a natural location for the island now presents itself, as marked by the red dot in the figure. Extremal surfaces that reach this $ z =\epsilon$ location can have finite, non-divergent areas. In a more conventional picture, these surfaces will look schematically like shown in Fig. \ref{funnel}, which meets the expectations from \cite{Mahajan2}. The key point is that the $x$-$y$ coordinates which are useful for exhibiting the structure of the black funnel have the property that the entire $x=0$ hyperbolic black hole collapses on to the boundary in the Fefferman-Graham  frame.  \begin{figure}[H]\centering
	\includegraphics[height=70mm,width=90mm,trim = {0 5cm 0 9cm}, clip]{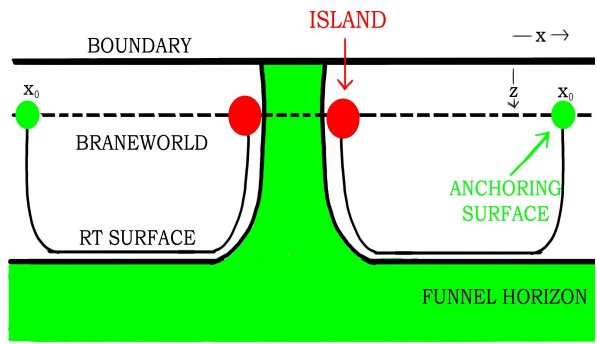} \\
	\vspace{-2cm}
	\caption{Black funnel with candidate islands and RT surface. }
	\label{funnel}
\end{figure} 
\noindent In AdS$_3$ we will treat the collapsed neck as the  analog the hyperbolic black hole, and therefore in order to model the higher dimensional examples, we will look at branes that cut the bulk (green) horizon near the $x=0$ region in Fig. \ref{AdS3}, similar to what happens in the higher dimensional Fig. \ref{rectangle}.
 
\end{itemize}

A key question we will be unable to address in this paper is that of explicit construction of such extremal surfaces in higher dimensions. Since the metric of \cite{Santos-Way} is only known numerically, this problem is best suited for those with the appropriate skills. Our motivation for believing their existence is threefold. Firstly, the structure of these solutions on a given timeslice, in light of our discussions in this section, is strikingly parallel to the sub-critical case considered in \cite{Mahajan3}. Indeed, one of our motivations to consider black funnels was that they naturally geometrize the intuition of \cite{Mahajan3} in a critical braneworld context. Secondly, in the next section, in a three dimensional funnel, we will be able to do very closely related calculations explicitly and identify the second extremal surfaces. We will present a discussion on AdS$_4$ funnels in hyperbolic backgrounds in \cite{Budha}. Thirdly, we will see that there is a simple and suggestive way to incorporate the cosmological nature of the braneworld into these discussions, to include islands.  

To directly check these claims in higher dimensions, we will need to take three steps. Firstly, we need explicit black funnel metrics in higher dimensional AdS. The state of the art in this is the work of Santos and Way \cite{Santos-Way} and the results are numerical. Secondly, we will need to construct explicit numerical RSII braneworld black holes at finite cut-off, induced from these AdS black funnels. This will loosely follow the structure of \cite{Figueras}, but note that the ``boundary" condition in the deep bulk that should be used for constructing this solution is not that of \cite{Figueras}, but that of a planar black hole. The braneworld will be time-dependent. This existence step is necessary, if we want to be fully confident that the claims that are being made by the extremal surface calculations in the next step, are indeed saying something about the information problem. Thirdly, we should construct extremal surfaces anchored on the brane in analogy with \cite{Mahajan3}, again numerically. As every step requires numerics, it is clear that this is a problem tailor-made for numerical relativists. 

Once one convinces oneself that the second extremal surfaces exists in black funnel backgrounds, that resolves (modulo a caveat we will expand on in section 5) an eternal black hole version of the information paradox, following the logic of \cite{Mahajan2, Mahajan3}. It would of course be interesting to see if one can make such an argument in the context of an evaporating black hole as well. A doubly-holographic calculation of this type may be doable in a dynamical black droplet geometry \cite{Hubeny, Hubeny2}. 


\section{Islands and Funnels in AdS$_3$}


In this section, we will specialize the discussion to the case of an AdS$_3$ black funnel presented in \cite{Hubeny}. This is a simple context where an analytic funnel metric is known. The result of our calculation will demonstrate the existence of the kind of extremal surface in this geometry that has a chance of leading to islands in higher dimensions. 
 
One difference between the higher dimensional settings of the last section and the present one, is that the collapsed neck of the AdS$_3$ funnel (which has some similarities to the zero energy hyperbolic black hole in higher dimensions) extends a finite distance from the boundary. A related point is that if one wants to take our calculation as a statement about a 1+1 dimensional information paradox, one should embed a brane in the 2+1 dimensional AdS funnel geometry. This brane can be nearby and parallel to the $z=0$ boundary\footnote{We will momentarily discuss the metric and coordinates of the geometry in more detail.} in the boundary asymptotic directions ($x \rightarrow \infty$), but will have to cut the $z=\sqrt{2}$ end of the green horizon in the figure in order for it to be truly analogous to the higher dimensional discussions (see the brane in Fig. \ref{rectangle}). It is conceivable that the behavior of such a brane at late times, especially near the bulk horizon, needs a more detailed analysis.
\begin{figure}[H]\centering
	\includegraphics[height=70mm,width=90mm,trim = {0 5cm 0 9cm}, clip]{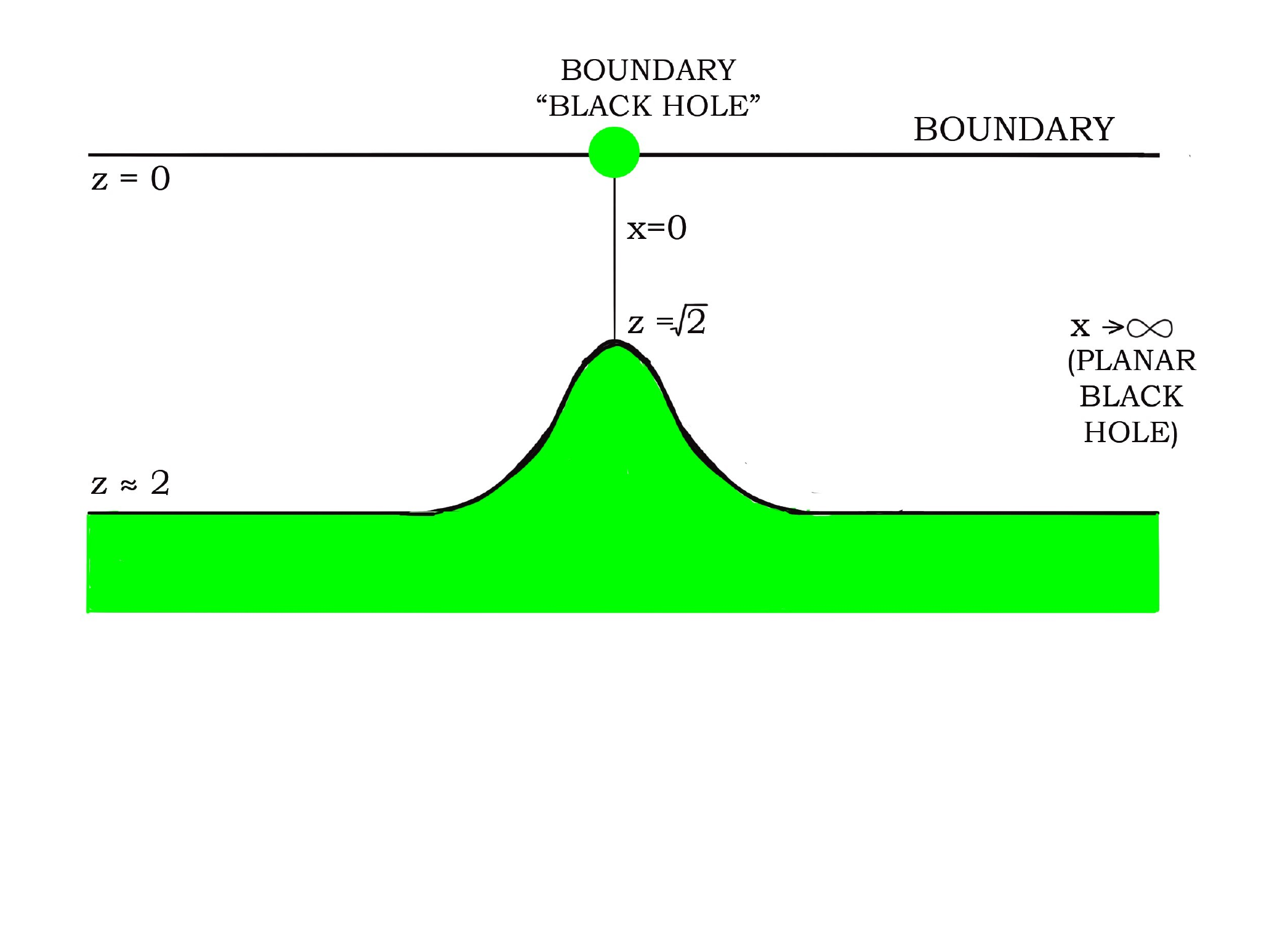} \\
	\vspace{-2cm}
	\caption{Spatial slice of the AdS$_3$ black funnel.  }
	\label{AdS3}
\end{figure} 
\noindent 
Another comment is that a 1+1 dimensional brane needs to have a gravity action on it (as in the JT gravity examples in \cite{Mahajan1}), but in higher dimensions one can simply work with the induced gravity on the brane. These things are important for various purposes, but they are details for us, because our goal here will be to simply demonstrate the existence of boundary-anchored bulk RT surfaces that straddle the bulk horizon. It is the existence of the latter that we expect, will generalize to higher dimensions. 

In other words, the parallel we are making is between the collapsed horizon of the AdS$_3$ funnel neck and the zero energy hyperbolic black hole in higher dimensions (which we discussed in the previous section). In higher dimensions, the bulk horizon goes all the way to the boundary, but we expect the analogy with the evanescent horizon to still hold. See in particular the coordinates illustrated in Fig. \ref{rectangle}. These are the natural coordinates to see the hyperbolic black hole, and they bring out the parallel with the AdS$_3$ case. It is near the location where the hyperbolic black hole meets the bulk horizon that we can expect to find extremal surfaces in higher dimensions. In analogy, we look for extremal surfaces in AdS$_3$ where the evanescent neck horizon  meets the bulk horizon. Finding them will be viewed as evidence for  our general picture.


Let us make one more comment before proceeding. We will not discuss the AdS$_4$ black funnels of \cite{Hubeny} here for the reasons discussed already, but since this seems to be the only other setting where analytic black funnels with a non-compact asymptotic region are known, these solutions are perhaps of some interest anyway \cite{Budha}.

With these caveats emphasized, we turn to the AdS$_3$ black funnel \cite{Hubeny}. The metric is 
\bea
d s^{2}=\frac{1}{z^{2}}\left(-f(x, z) d t^{2}+g(x, z) d x^{2}+d z^{2}\right)
\eea
with 
\bea
f(x, z)=\tanh ^{2} x\left(1-z^{2} \frac{\cosh ^{2} x+1}{4 \cosh ^{2} x}\right)^{2}, \ \ 
g(x, z)=\left(1+z^{2} \frac{\cosh ^{2} x-3}{4 \cosh ^{2} x}\right)^{2}
\eea
where $x$ is a radial coordinate and $z$ is the holographic direction with boundary at $z=0$. A picture of a constant time slice of the spacetime can be found in Fig. \ref{AdS3}. The horizon is obtained by setting the metric function $f=0$, and it has one branch coming from $x=0$ corresponding to the $\tanh$ piece (this corresponds to the collapsed neck), and another that can solved explicitly for $z$ in terms of $x$ which yields the green horizon in the figure. Asymptotically as $x \rightarrow \infty$ the latter tends to the planar black hole horizon at $z \sim 2$. It deviates from the planar black hole near $x \sim 0$ and we get the bump in the figure which is peaked at $z=\sqrt{2}$.  

Based on our higher dimensional prejudices discussed in the last section, we expect that there should be an RT surface close to this bump that reaches out to the asymptotic region of the boundary $z=0$ as $x \rightarrow \infty$. If we put a brane in the geometry analogous to the one in Fig. \ref{rectangle} (like we discussed above Fig. \ref{AdS3}), we are to look for an extremal surface anchored on the brane between a fixed large $x=x_0$ and an $x=x_h \sim 0$, and pick the minimal one by varying over $x_h$. An analogue of this was done in \cite{Mahajan3}. Since we wish to avoid the complication of explicitly introducing the brane, we will instead follow a path following the spirit of \cite{Geng} (see also \cite{Myers}). We will look for extremal surfaces anchored to $x=0$\footnote{We emphasize again that when a brane is introduced, this will have to be slightly modified.} and $x=x_0$, with $dz/dx|_{x=0}=0$. We expect that the extremal surface will be normal to the (would be) brane, in the limiting case where the hanging brane coincides with the evanescent neck. We expect the system to be symmetric with respect to a quotienting across the neck. This is loosely parallel to the orbifolding argument used across the non-critical brane in \cite{Geng}. If we were to consider more general brane configurations, we would have to be more careful, but our goal is merely to make an existence argument for horizon-straddling extremal surfaces in this geometry. As we will see momentarily, the extremal surfaces that we find naturally have features one expects from island-like configurations. 

\begin{figure}[H]\centering
	\includegraphics[angle=360,origin=c]{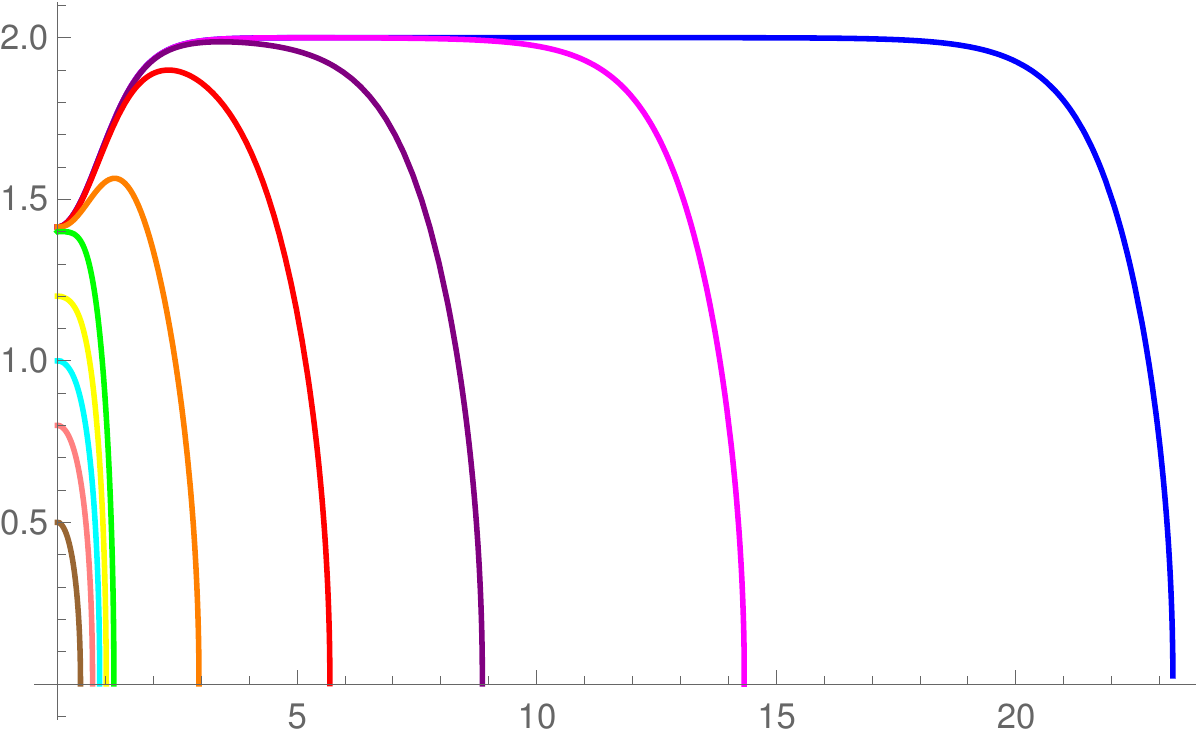} \\
	\caption{Plots of extremal surfaces with $x=0$ and $dz/dx|_{x=0}=0$. The vertical axis is $z$ and the horizontal axis is $x$. The ``island" extremal surfaces are the ones that straddle the horizon, note that the bump in the horizon as shown in Fig. \ref{AdS3} happens in this plot at $(x=0, z=\sqrt{2})$.}
	\label{Plot}
\end{figure}

The relevant integral that we have to extremize  then is 
\bea
I=\int_{x=0}^{x=x_0} \frac{dx}{z}\sqrt{g(x,z)+\left(\frac{dz}{dx}\right)^2}
\eea
with the boundary conditions that $dz/dx|_{x=0}=0$ and $z(x=x_0)=0$.  Note that $I$ has dependence on both the independent variable $x$ as well as $z$ and $dz/dx$. The variational equation of motion arising from this integral is an ODE, but it is too complicated to be instructive (as far as we could see). So we will not exhibit it. Solving it for fixed values of $x_0$ numerically is possible by using a shooting method, but what we will do is to integrate the Euler-Lagrange equation from $x=0$ with $dz/dx|_{x=0}=0$ for various values of $z(x=0)$. From the plots of such curves, it will become clear that any value of $x_0$ can be arrived at if one chooses a suitable $z(x=0)$, establishing the existence of the extremal surface. The task is easy enough for simple numerical integrators of ODEs (we used Mathematica) to do, and the results are presented in Fig. \ref{Plot}.  
\noindent
The evidence for island like physics is clear from the plots. As the anchoring point at $x=0$ gets closer to the horizon, the second anchoring point gets pushed out to larger and larger $x_0$. The extremal surfaces straddle the green horizon for longer and longer before diverging from it and hitting the boundary/brane. Let us also note that if there were a brane like the one in Fig. \ref{rectangle} in this geometry, the island would get closer and closer to the braneworld horizon, as the anchoring point $x_0$ at the boundary/brane gets larger and larger. This is again consistent with the observations of \cite{Mahajan2} and is also intuitive. The crowding of the islands near $x=0$ as a function of $x_0$ is also worth noting. 

The above discussion takes care of the ``second" extremal surfaces that we defined earlier, in analogy with \cite{Mahajan3}. Now we turn to the ``first" kind of extremal surfaces, those that cut the planar black hole horizon. Here it turns out that it is beneficial to write the integral that captures the area in a notation where $z$ is the independent variable:
\bea
I'=\int_{z=0}^{z=z_*} \frac{dz}{z} \sqrt{1+g(x,z)\left(\frac{dx}{dz}\right)^2}  
\eea 
Here $z_*$ is the location on the horizon where $dx/dz =0$\footnote{At least for large enough $x_0$, we can expect this to be true.}, and $x(z=0)=x_0$. The usefulness of this form is that the Euler-Lagrange equation has the property that when $x'=0$, it reduces to $x''=0$. In other words, if we are imagining starting the integration from the horizon where $dx/dz =0$, we can see that the extremal surface is simply $x= x_0$.  This is the first kind of extremal surface.

Qualitatively similar comments apply to general dimensions. We expect the structure of the extremal surfaces in higher dimensions to be superficially slightly different from the ones we constructed in AdS$_3$, as indicated in Fig. \ref{funnel}. In particular we cannot put the boundary conditions on the (analogue of) $x=0$ as we did here, because the hyperbolic black hole is repulsive to RT surfaces, and more directly because in the Fefferman-Graham gauge it has collapsed to the boundary. But we expect anchoring them as shown in Fig. \ref{funnel}, which is what one would expect from an island prescription, to be possible. Note that it is precisely this structure that one expects from the form of the branes in Figs. \ref{triangle} and \ref{rectangle}. Note that the latter structure is parallel to the one we found in 3 dimensions. Our observations suggest that any brane that is asymptotically parallel to the boundary and also cuts the bulk horizon, may allow islands very similar to the ones we have discussed.

\section{Cosmological Islands}

The existence of these extremal surfaces on the constant time slices of the metric, as we saw above, is non-trivial. Furthermore, it is natural that for large enough $x_0$, the second kind of extremal surfaces have bigger area at some initial time than the first kind -- the former scale extensively with $x_0$, the latter do not. Also  as time elapses, the contribution from inside the horizon to the area of the first extremal surface can be expected to increase linearly with time. Because of all these reasons, one might hope that there is a phase transition at late times, and that the argument for avoiding the information paradox is just a re-run of that in \cite{Mahajan3}. 

But there is one fly in the ointment -- the brane is moving in the bulk\footnote{
I thank A. Karch for emphasizing this to me.}. This means that the island argument here needs some adaptations from that in \cite{Mahajan3}. We will have to find a way to keep track of {\em only} the black hole contribution to the entanglement entropy, and this is the purpose of this section. Remarkably, we find that there is a quite compelling way to do this, both from the bulk and the brane perspectives. Since the argument here contains some interesting new ingredients that did not play a role in \cite{Mahajan3}, this can be viewed as additional evidence for the role of islands. Note also that the discussion in this section goes a bit beyond the purely kinematical discussion of extremal surfaces that we were concerned with in the rest of the paper. 

Even though the geometry for a full braneworld black funnel is not known, we know that it tends to a planar black hole far away from the funnel. A brane in a planar black hole has to accelerate, so that it does not fall through the horizon. On the worldvolume of the brane this results in an FRW cosmology, with the scale factor directly captured by the radial position of the brane \cite{Kraus}. Even though the braneworld attached to a black funnel is likely hard to (numerically) construct, a natural guess is that the horizon region of the brane also moves, and therefore experiences the FRW scale factor\footnote{I thank P. Figueras, A. Karch and J. Santos for correspondence on this point.}. If it does not, it is easy to convince oneself that the arguments of \cite{Mahajan3} apply with minor changes here as well, and the information problem is solved already at this stage. But we believe the braneworld horizon stretches, and therefore the situation is more interesting. Note eg., that the AdS/CFT black funnel has a horizon on the conformal boundary. This seems more natural to understand in terms of the braneworld funnel if the horizon were to {\em not} get stuck at a finite location in the AdS radial direction. 

Now a moving brane with a cosmology on it makes our problem  different from that in \cite{Mahajan3}. So let us first observe that the eternal black hole information paradox in this braneworld system is still well-posed. The entire system is getting red-shifted and cooling down due to cosmological expansion, but the temperature of the black hole remains the same as that of radiation and therefore it remains in equilibrium\footnote{Note that there is also a cosmological piece in the entanglement entropy, so the paradox is best phrased and resolved in comoving coordinates as we will see below.}. By picking the confinement scale of the matter and the initial size of the black hole suitably, we can easily ensure that we stay above the deconfinement temperature for order of a few Page times. The relevant Page time \cite{Mahajan2} is given by $\sim \beta S_{BH}/c$ where $c$ is the large central charge of the matter. Therefore the paradox can still be arranged to exist, by choosing parameters. If we pick the black hole to be small enough (but still macroscopic\footnote{Let us note that primordial-sized black holes have been argued to exist upto the Planck limit of $10^{-8}$ kg \cite{Hawking2}, but we do not need anywhere near that extreme. To give a rough idea -- a black hole of mass 1 kg is sufficiently macroscopic, but will have a Hawking temperature of $ \sim 10^{23}$ K, which is easily higher than the deconfinement temperature of real world QCD, which is about $\sim 10^{12}$ K. Note that the (large) central charge and the deconfinement scale of matter, we will treat as free inputs.}) with a Page time that is small enough, the Universe need not cool too much during that timescale. The fact that smaller black holes have higher temperature but lower entropy, works in our favor here. The matter on the brane is deconfined radiation and the scale factor goes as a power law $a(t)\sim t^{1/2}$ for a 4D braneworld (and temperature falls as $\sim 1/a$), even though the specific time dependence of the scale factor does not affect our discussions below. 

A second point worthy of note is that because of the FRW scale factor, there are two natural candidates for the entanglement entropy that one can compute on the braneworld. The first possibility is that we compute the entanglement entropy of a sphere located at fixed physical distance from the braneworld black hole. The other is that we look at a sphere that is at a fixed comoving radius. Together with this choice, we also have to make a corresponding choice for the UV cut-off in physical distance or comoving distance. 

We will argue that since our goal is to keep track of the entanglement entropy due to the black hole and not due to the cosmological expansion, it is the comoving entanglement entropy that one should compute. Let us present a few reasons why think this is an eminently natural choice. 
\begin{itemize}
\item From the point of view of an entanglement entropy computation on the brane, note that if we fix a comoving length scale as the UV cut-off, we will never trace over the modes that become accessible due to cosmological red-shifting. This is precisely what we would like, since we wish to avoid counting the cosmological contribution to the entanglement entropy and only keep track of the black hole physics.
\item Note that this statement can be understood also from the bulk/brane geometric point of view. The time-dependence is a direct result of embedding the brane in the bulk planar black hole geometry, and on the brane it gets reflected in the fact that the black hole is living in a cosmology. The key point is that even when there is no funnel or braneworld black hole, but only the bulk planar black hole, we get cosmology/time-dependence on the brane. What the comoving approach does is to regulate this contribution away. 
\item A practical reason for preferring comoving coordinates is that since the brane horizon is stretching, it will reach any fixed physical radius in finite time. It may perhaps be possible to choose one's scales so that this happens after the Page time, but it certainly does not look too appealing as a starting point.
\item Another way to see that the natural choice is the comoving location, is to look at the bulk picture of the extremal surfaces. In Fig. \ref{funnel} the brane is moving ``up" because it is cosmological, and fixed locations in boundary coordinates to anchor the extremal surfaces naturally translate to fixed comoving coordinates on the braneworld. On the contrary, if we were to anchor the extremal surfaces at fixed physical radius, as the brane evolves, they will have to move inward in the $x$-direction in the figure. This is a bulk manifestation of the previous bullet point.
\item Note that the extremal surfaces familiar from AdS/CFT, naturally map on to {\em comoving Ryu-Takayanagi surfaces} from the brane point of view. This is evident from the structure of cosmological branes in general \cite{Kraus}, as well as from the structure of Figs. \ref{funnel}, \ref{AdS3} and \ref{Plot} (note that the scale factor on the brane is a direct result of the brane moving ``up").
\end{itemize}

A crucial observation is that if we are computing entanglement entropy of regions defined by their comoving coordinates, and with a fixed comoving cut-off, the scale factor drops out of the entanglement entropy and therefore its leading behavior is time-independent. This is {\em not} the case if we fix a physical cut-off instead. Note again that this is consistent with our previous expectation that the comoving calculation is insensitive to cosmological production of entanglement entropy. Finally, let us also emphasize that since RT surfaces anchored on the brane are of manifestly finite area, the dual entanglement entropy defined on the brane should necessarily be understood with a cut-off.

Now we turn to our bulk computation of entanglement entropy using the RT surfaces and how they should be interpreted. An immediate observation is that because the branes are moving ``up", the contribution coming from near the anchoring surfaces has time dependence\footnote{One might hope that this will go away in the subtraction with respect to the trivial (``first") RT surface, if ones goal was to simply try to reproduce the methods of \cite{Mahajan3}. But it is clear that the island contribution also has the time dependence and that will not go away in the subtraction. In any event, as we will argue, this time dependence has a physical origin and hoping that subtraction would save us, is unsatisfactory. Note in particular that these are manifestly finite quantities, and therefore they should be explicitly understood with a cut-off from the brane perspective.}. Our discussion in the last paragraph is a strong indication that this time dependence is a consequence of implicitly working with a {\em physical} UV cut-off.  Further evidence for this comes by a direct calculation -- we can estimate how the RT surface area depends on the (moving brane) location $r$. Here $r$ is the Poincare radial coordinate at the anchoring surface on the brane. Note that the scale factor on the brane is determined by the time dependence of its radial position \cite{Kraus}, which means that we should eventually take $r \sim a(t)$. In $(d+1)$-dimensional AdS, the cut-off dependence of RT area is well-known and gives rise to the area law for the entanglement entropy on the boundary\footnote{It can also be estimated by a direct calculation, noting that near the cut-off in AdS, the RT surface will have the metric $dr^2/r^2+r^2 (dX^2)$ where $dX^2$ denotes some $r$-independent, $(d-2)$-dimensional geometry. This leads to an area (which is the volume of the RT surface) that can be estimated as $\int \frac{dr}{r} r^{d-2} \sim r^{d-2}$.}, and therefore should scale as $r^{d-2}$. For a moving brane, with the brane position $r \sim a(t)$  this therefore gives precisely the scale factor dependence one expects due to a physical UV cut-off in the braneworld theory. And the remedy is clear - one has to change the cut-off to a comoving cut-off, and  all one has to do to accomplish this at leading order is to delete the scale factor dependence arising at the anchoring surfaces in our RT calculation. Essentially, the object we are calculating is 
\bea
S \sim \frac{A(t)}{\epsilon^{d-2}(t)} \sim \frac{a(t)^{d-2}}{\epsilon^{d-2}(t)}
\eea 
where $A$ is a suitable area on the brane. There are two natural choices for the UV cut-off $\epsilon(t)$. The physical cutoff takes $\epsilon(t) = \epsilon_{phys}$, a constant length independent of time. The comoving cutoff instead fixes the coordinate grid length as the fundamental quantity, and sets $\epsilon(t) \sim a(t) \epsilon_{comov}$ where $\epsilon_{comov}$ is a constant cutoff in the comoving grid size. The key point is that to go between the two definitions of $S$ one needs to scale by a factor of $a(t)^{d-2}$, and the key observation is that it is precisely this factor that shows up as the time-dependence of the RT surfaces at the anchoring surfaces on the brane.  

Of course, since the full geometry of the braneworld black funnel is a complicated time-dependent geometry, evaluating these explicitly will require a computer. But by identifying the relevant object as the entanglement entropy with a comoving cut-off, we have understood how the superficially time-dependent RT area leads ultimately to a meaningful time-independent result. And it is at this stage, that we are free to blissfully apply the argument of \cite{Mahajan3}. We note that second RT surface area (when computed correctly with the comoving regulator) is time independent, and therefore there will indeed be a phase transition at late times. This is a very strong suggestion that cosmological islands resolve a version of the information paradox for these eternal cosmological black holes.


Because the branes are at finite cut-off, all our extremal surface calculations (when interpreted as brane calculations) give manifestly finite answers\footnote{Note that moving from the physical cut-off to the comoving cut-off was a finite (albeit time-dependent) renormalization.}. 
Because they are critical braneworlds, the gravitons in them  are massless as long as we are in sufficiently high dimensions. This means in particular that the problem raised in \cite{Geng} does not arise here.

\section{Conclusion}

Let us first summarize the context of this paper. When JT-gravity is coupled to a sink CFT in 2D, the emergence of the Page curve can be understood via semi-classical islands in a 3D ``doubly-holographic" geometry. To this end, one views JT-gravity as living on a Karch-Randall braneworld. The higher dimensional analogues of this set up lead to black holes in AdS braneworlds, but with massive gravitons. It has been observed that in the massless limit of these massive gravitons, islands also vanish, raising questions about the validity of the prescription for real world gravity. 

In this paper, we noted that by working with a critical Randall-Sundrum II set up instead of Karch-Randall, we can get braneworlds with vanishing cosmological constants and massless gravitons. Stringing together various known facts in the literature (some from numerical relativity), we argued that a version of the doubly-holographic Ryu-Takayanagi calculation should make sense in the critical RSII braneworld, in terms of a black funnel attached to the braneworld black hole. The geometry of this configuration makes it plausible that islands exist. We checked this by direct calculation in a (not fully realistic) 3-dimensional example, where the analytic funnel metric is explicitly known. This could resolve an eternal black hole version of the information paradox, in a suitable braneworld in this system. Along the way, we understood a thing or two about the role of islands in cosmology -- in particular the usefulness of the comoving cut-off on the brane.

Note that critical braneworlds are interesting for two reasons. Firstly, they provide a doubly holographic geometry for non-AdS black holes. Secondly, they give us massless gravitons. The arguments we present here are complementary to those in \cite{KPP}, and further strengthen the case that islands may be of more general significance than 1+1 dimensions or AdS, in resolving the Page curve version of the information paradox. 

One interpretation of the recent developments on the information paradox is that the entropy problem in Hawking's calculation can be solved already at the semi-classical level. What we mean by this is that the exact (as in not coarse-grained) entropy can be computed with semi-classical gravity, without knowledge of the UV theory. This is surprising. On the other hand, a full understanding of the microstate of the system, will require more work. Doubly holographic (indeed any semi-classical) calculation assumes that there is some UV completion to which these calculations can be viewed as approximations to. The outstanding question, as always, is to understand this UV completion better.

\section{Acknowledgments}

I thank Budhaditya Bhattacharjee for discussions, and also for making the pdf versions of Figures 1 through 5. I also thank Pau Figueras, Jorge Santos and especially Andreas Karch for helpful correspondence. 

\appendix

\end{document}